\documentclass[10pt,conference]{IEEEtran}
\usepackage{epsfig,rotating,setspace,latexsym,amsmath,epsf,amssymb,amsfonts,bm,theorem,cite,authblk, bbm,color, hyperref, multirow, caption, subcaption}
\IEEEoverridecommandlockouts
\allowdisplaybreaks

\title{Age of Information in Deep Learning-Driven Task-Oriented Communications \thanks{This work was supported in part by the Commonwealth Cyber Initiative (CCI), an investment in the advancement of cyber R\&D, innovation, and workforce development.}}
\begin{document}
\author[1]{Yalin E. Sagduyu}
\author[2]{Sennur Ulukus}
\author[3]{Aylin Yener}

\affil[1]{\normalsize  Virginia Tech, Arlington, VA, USA}

\affil[2]{\normalsize University of Maryland, College Park, MD, USA}

\affil[3]{\normalsize  The Ohio State University, Columbus, OH, USA}
\maketitle
\begin{abstract}
This paper studies the notion of age in task-oriented communications that aims to execute a task at a receiver utilizing the data at its transmitter. The transmitter-receiver operations are modeled as an encoder-decoder pair that is jointly trained while considering channel effects. The encoder converts data samples into feature vectors of small dimension and transmits them with a small number of channel uses thereby reducing the number of transmissions and latency. Instead of reconstructing input samples, the decoder performs a task, e.g., classification, on the received signals. Applying different deep neural networks of encoder-decoder pairs on MNIST and CIFAR-10 image datasets, the classifier accuracy is shown to increase with the number of channel uses at the expense of longer service time. The peak age of task information (PAoTI) is introduced to analyze this accuracy-latency tradeoff when the age grows unless a received signal is classified correctly. By incorporating channel and traffic effects, design guidelines are obtained for task-oriented communications by characterizing how the PAoTI first decreases and then increases with the number of channel uses. A dynamic update mechanism is presented to adapt the number of channel uses to channel and traffic conditions, and reduce the PAoTI in task-oriented communications.
\end{abstract}


\section{Introduction} \label{sec:Intro} 
In \emph{conventional communications}, the goal is to optimize information transfer by accounting for channel impairments. The transmitter and receiver operations can be individually or jointly designed to reconstruct the information transferred from a transmitter to its receiver by minimizing a reconstruction loss such as the symbol error rate. For that purpose, \emph{deep neural networks} (DNNs) can be used to model the transmitter and receiver operations such as channel coding/decoding and modulation/demodulation jointly as an \emph{autoencoder} to minimize the end-to-end reconstruction loss.

The goal of communications has been extended to preserve the meaning of information in \emph{semantic communications} \cite{guler2014semantic, gunduz2022beyond, uysal2021semantic}. For that purpose, the loss for training the autoencoder can incorporate both the reconstruction loss in conventional communications and the semantic loss, i.e., the loss of meaning during the information transfer \cite{Semanticadversarial, sagduyu2022vulnerabilities}. Semantic communications has been considered to preserve the meaning in information transfer of different data modalities, including  \emph{text} \cite{guler2018semantic, xie2021deep}, \emph{image} \cite{qin2021semantic, Semanticadversarial}, \emph{video} \cite{Geoffreyvideo}, and \emph{speech/audio} \cite{weng2021semantic, walidaudio}. 

The semantics of information can also represent the significance of information relative to a task that necessitates the information transfer (e.g., edge sensor devices take images and the fusion center needs to detect intruders in these images). To that end, \emph{task-oriented communications} or \emph{goal-oriented communications} \cite{shao2021learning, strinati20216g, kang2022task} changes the paradigm of conventional communications in a way that the objective is not anymore reliable information reconstruction, but it is the successful execution of a task (e.g., a classification task) at the receiver while the data is available at the transmitter. The transmitter operations such as source coding, channel coding, and modulation are modeled as an encoder to generate and transmit feature vectors of small dimension, whereas the receiver does not run the usual receiver chain but directly employs a decoder to perform the task such as classifying the received signals without reconstructing input samples such that only a small number of transmissions is needed with low latency. By accounting for both channel and data characteristics, the \emph{encoder-decoder pair} is trained jointly as an end-to-end DNN to optimize the task performance such as the categorical cross-entropy for the classification loss \cite{TOCattacksagduyu}. 

In task-oriented communications, the task execution can be considered a \emph{status update} when data samples arrive randomly at the transmitter queue. Then, the performance can be measured not only by accuracy but also by \emph{information timeliness}. To that end, the study of the \emph{age of information} (AoI) has been instrumental to characterize how the information ages over time in a queuing system where a transmitter aims to send timely status updates to its receiver \cite{kaul2012real}. The peak AoI (PAoI) has been considered a more tractable metric for information timeliness \cite{PAoI}. The AoI metrics have been considered when executing a task such as real-time source reconstruction  \cite{pappas2021goal} or computation offloading at the edge \cite{timely-dist-comp, qin2022timeliness}.  Since the AoI metrics do not account for the information dynamics at the source, the binary freshness metric has been employed in \cite{bastopcu2020information} to compare the information at the receiver with the information stored at the transmitter. To measure the freshness of informative updates, the Age of Incorrect Information (AoII) has been proposed in \cite{maatouk2020age} by combining the AoI and conventional error penalty functions. By considering status updates subject to a distortion, a trade-off has been identified in \cite{bastopcu2021age} between the quality of information and the AoI (processing longer at the transmitter reduces the distortion of the update but it also increases the age); see also partial updates \cite{bastopcu2020partial}. Recently, supervised and reinforcement learning approaches have been employed for keeping information freshness \cite{Leng-Yener-22}.

In this paper, we consider the timeliness in task-oriented communications, when a machine learning task needs to be performed at the receiver without a need for information reconstruction. A motivating example is an inter-vehicle or drone network, where each node (vehicle or drone) takes images and exchanges them with its neighbor nodes. Instead of transmitting the entire image data, each node transmits only a reduced amount of information such that the receiver node can still correctly classify images to labels (e.g., traffic signs or targets). It is important to complete the image classification task in both correct and timely manner. Thus, we consider the age of status updates corresponding to correct task completion and jointly optimize the information transfer and classification operations by training an \emph{encoder-decoder pair}. Both the classification accuracy and the latency (service time) increase with the \emph{number of channel uses} (i.e., the size of the encoder output or the size of the decoder input) that emerges as an important design parameter in task-oriented communications. 

We consider \emph{image classification} as the task, use MNIST and CIFAR-10 datasets, and train different feedforward (FNN) and convolutional neural network (CNN) models for the encoder-decoder pair in task-oriented communications. First, we characterize classification accuracy as an increasing function of the number of channel uses and the signal-to-noise ratio (SNR). Then, we derive the Peak Age of Task Information (PAoTI) as a function of the number of channel uses, the SNR, and the arrival rate assuming that age continues to grow unless a successful task execution, i.e., correct image classification, takes place as a status update. We show that the PAoTI decreases first and then increases with the number of channel uses. This identifies the  number of channel uses to minimize the PAoTI as a design feature for task-oriented communications. When the SNR and the arrival rate are not known in advance, we present a \emph{dynamic update} mechanism that adapts the number of channel uses to arrival and channel conditions over time and reduces the PAoTI. The results highlight the accuracy-latency tradeoffs in terms of the PAoTI and identify new design guidelines for task-oriented communications.

The rest of the paper is organized as follows. Section \ref{sec:TOC} describes the task-oriented communications system based on deep learning and evaluates the classification accuracy. Section \ref{sec:PAoTI} analyzes the PAoI in task-oriented communications. Section \ref{sec:perf} evaluates the PAoI performance in task-oriented communications as a function of system parameters including the number of channel uses and presents a dynamic update of the number of channel uses. Section \ref{sec:Conclusion} concludes the paper. 

\section{Deep Learning-Driven Task-Oriented Communications} \label{sec:TOC}
\begin{figure}[t]
\centering
\includegraphics[width=\columnwidth]{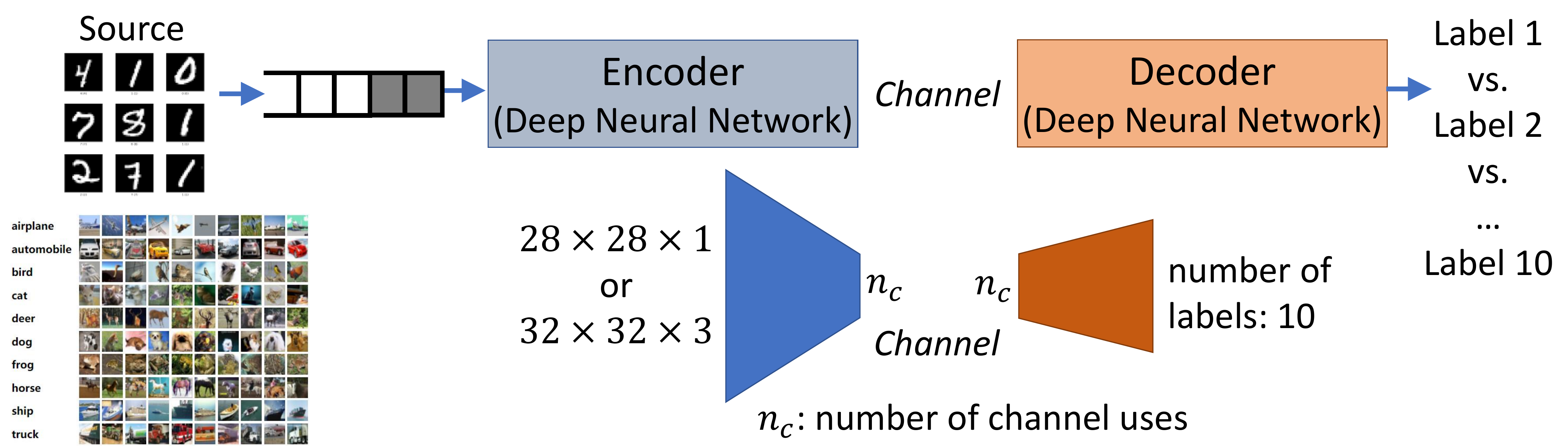}
 \caption{Task-oriented communications.} 
 \label{fig:AOI_TOC_system}
 \vspace*{-0.4cm}
\end{figure}

We consider task-oriented communications driven by deep learning. As shown in Fig.~\ref{fig:AOI_TOC_system}, the transmitter and the receiver operations are represented as an encoder and a decoder, respectively, and they are jointly trained. The data samples such as images are the input to the encoder. The encoder incorporates the operations of source coding, channel coding, and modulation, and converts the input sample to modulated signals. The size of the output of the encoder is smaller than the size of the input sample, i.e., the encoder captures lower-dimensional latent features that are transmitted over the channel with a small number of channel uses.  

The signals received at the receiver side are given as input to the decoder that executes a task, namely classifies received signals to the labels of input data samples at the transmitter. The encoder and decoder are jointly trained as an end-to-end deep neural network while accounting for the channel. This setting is different from autoencoder communications \cite{Oshea1} that typically processes symbols (bits) as input at the transmitter and reconstructs them at the receiver, i.e., does not include source coding and decoding operations. Additionally, in our setting, the input samples are not reconstructed at the receiver. Since only feature vectors of a small dimension are transmitted over a limited number of channel uses, task-oriented communications is more energy-efficient and is better at correct classification (compared to data reconstruction followed by classification), as shown before for the task of spectrum sensing and wireless signal classification in \cite{TOCattacksagduyu}. 

We consider two image datasets:
\begin{itemize}
\item \textbf{MNIST:} The MNIST dataset is composed of images of handwritten digits \cite{MNIST}. The label of each data sample is its digit (from $0$ to $9$). There are total of 10 labels. Each data sample (image) is of $28\times28$ grayscale pixels and the value of each pixel is between $0$ and $255$. We consider different models: (a) the FNN where each data sample is represented by a feature vector of size $28 \times 28 = 784$, (b) the CNN where each data sample is of size $28 \times 28 \times 1$. The dataset consists of 60,000 training samples and 10,000 test samples. 
\item \textbf{CIFAR-10:} The CIFAR-10 dataset consists of color images from 10 classes, `airplane',	`automobile', `bird', `cat', `deer', `dog', `frog', `horse',	`ship', and	`truck \cite{CIFAR}. There are total of 10 labels. Each data sample (image) is of $32 \times32 \times 3$ color (RGB) pixels and the value of a given pixel in each red, green and blue component is between $0$ and $255$. We consider the CNN as the model to train (the FNN is known to have poor performance for the CIFAR-10 dataset). The dataset consists of 50,000 training samples and 10,000 test samples. 
\end{itemize}

In each case, the corresponding feature is normalized to $[0,1]$ and given as input to the encoder of the transmitter. The encoder that is run on each input sample reduces the dimension to $n_c$, which is defined as the number of channel uses to transmit the modulated symbols at the output of the transmitter (assuming one symbol can be transmitted in each channel use). The signal at the output of the encoder is transmitted with $n_c$ channel uses over an AWGN channel. The received signal (of dimension $n_c$) is given as input to the decoder at the receiver. The decoder's output is the classification label. Note that the input sample is not reconstructed at the receiver compared to conventional communications. Table~\ref{table:DNN} shows the architectures of the encoder and decoder for different data and model types, namely (a) MNIST + FNN, (b) MNIST + CNN, and (c) CIFAR-10 + CNN. Categorical cross-entropy is used as the loss function and Adam is used as the optimizer. The numerical results are obtained in Python and the models are trained in Keras with the TensorFlow backend. A Gaussian noise layer with the corresponding SNR is inserted between the encoder and the decoder to reflect the channel effects.
\begin{table}[h]
 \captionsetup{justification=centering}
     \caption{Encoder-decoder architectures for task-oriented communications.}   
    \label{table:DNN}
	\begin{center}
	\footnotesize
        \begin{subtable}[h]{0.45\textwidth}
         \caption{Data: MNIST, Model: FNN.}
		\begin{tabular}{l|l|l}
			Network & Layer & Properties \\ \hline \hline
			Encoder & Input & size: 28$\times$28$\times$1 \\
            & Dense &  size: $784$, activation: ReLU \\
			& Dense & size: $n_c$, activation: Linear \\ \hline
            Decoder & Input & size: $n_c$ \\ 
			& Dense & size: $n_c$, activation: ReLU \\
            & Dense & size: $10$, activation: Softmax \\
		\end{tabular}
        \end{subtable}
        
\vspace{0.2cm}

        \begin{subtable}[h]{0.45\textwidth}
            \caption{Data: MNIST, Model: CNN.}
		\begin{tabular}{l|l|l}
			Network & Layer & Properties \\ \hline \hline
			Encoder & Input & size: 28$\times$28$\times$1 \\
& Conv2D & filter size: 32, kernel size: (3,3) \\ & & activation: ReLU \\
& MaxPooling2D & pool size: (2,2) \\
& Conv2D & filter size: 64, kernel size: (3,3) \\ & & activation: ReLU \\
& MaxPooling2D & pool size: (2,2) \\
& Flatten & -- \\
& Dropout & dropout rate: 0.5 \\
& Dense & size: $n_c$, activation: Linear \\ \hline
Decoder & Input & size: $n_c$ \\
 & Dense & size: $n_c$, activation: ReLU \\
& Dense & size: 10, activation: Softmax \\
		\end{tabular}
        \end{subtable}
        
\vspace{0.2cm}

        \begin{subtable}[h]{0.45\textwidth}
        \caption{Data: CIFAR-10, Model: CNN.}
		\begin{tabular}{l|l|l}
			Network & Layer & Properties \\ \hline \hline
			Encoder & Input & size: 32$\times$32$\times$3 \\
& Conv2D & filter size: 32, kernel size: (3,3) \\ & & activation: ReLU \\
& Conv2D & filter size: 32, kernel size: (3,3) \\ & & activation: ReLU \\
& MaxPooling2D & pool size: (2,2) \\
& Dropout & dropout rate: 0.25 \\
& Conv2D & filter size: 64, kernel size: (3,3) \\ & & activation: ReLU \\
& Conv2D & filter size: 64, kernel size: (3,3) \\ & & activation: ReLU \\
& MaxPooling2D & pool size: (2,2) \\
& Dropout & dropout rate: 0.25 \\
& Flatten & -- \\
& Dense & size: 512, activation: ReLU \\
& Dropout & dropout rate: 0.25 \\
& Dense & size: $n_c$, activation: Linear \\ \hline
Decoder & Input & size: $n_c$ \\
 & Dense & size: $n_c$, activation: ReLU \\
& Dense & size: 10, activation: Softmax \\
		\end{tabular}
        \end{subtable}
        
	\end{center}
\end{table}

\begin{figure*}[h!]
\centering 
\begin{subfigure}[b]{0.32\textwidth}
\centering
\includegraphics[width=\columnwidth]{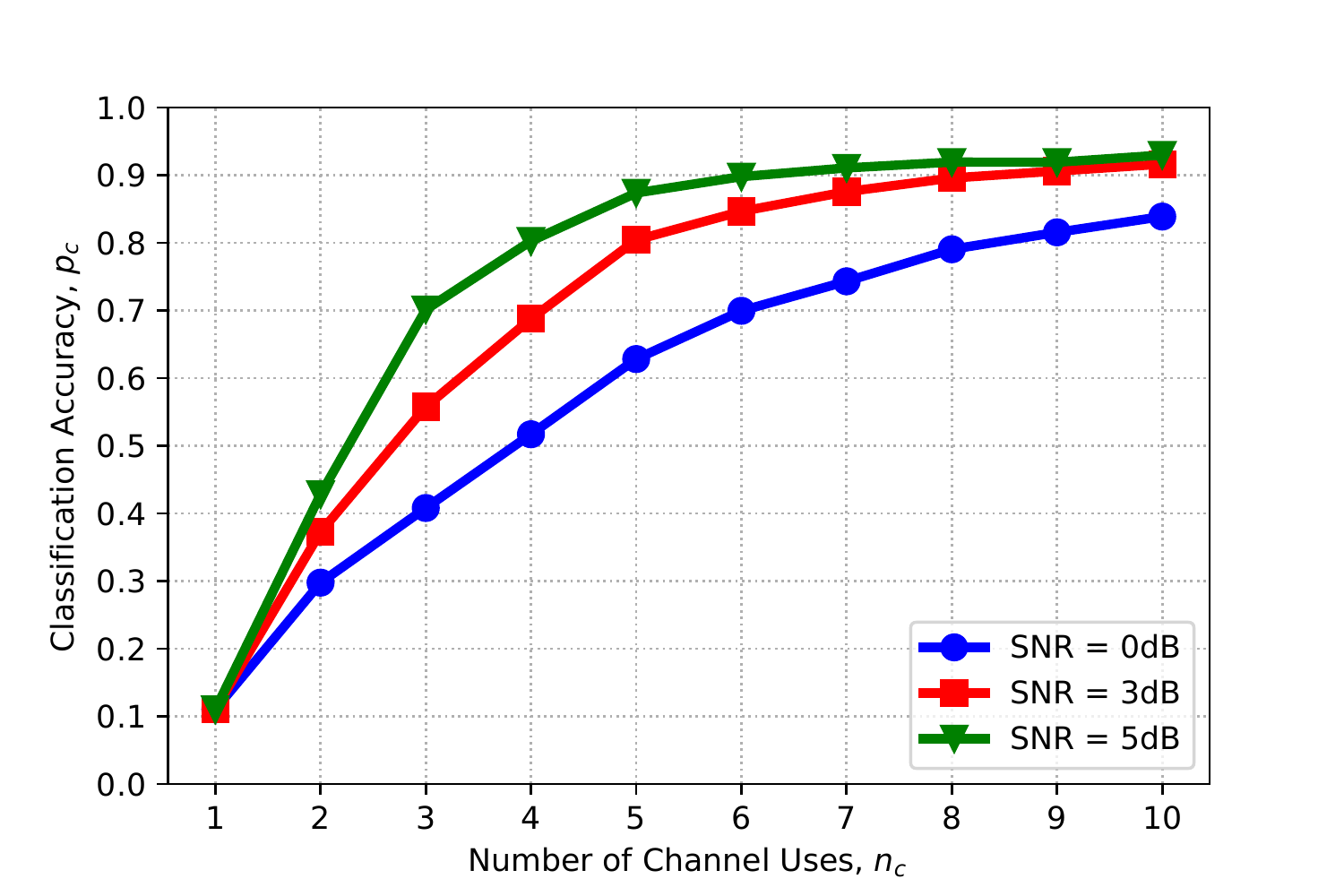}
\caption{Data: MNIST, Model: FNN.}
\label{fig:accuracyvsnchannelMNISTFNN}
\end{subfigure}
\begin{subfigure}[b]{0.32\textwidth}
\centering
\includegraphics[width=\columnwidth]{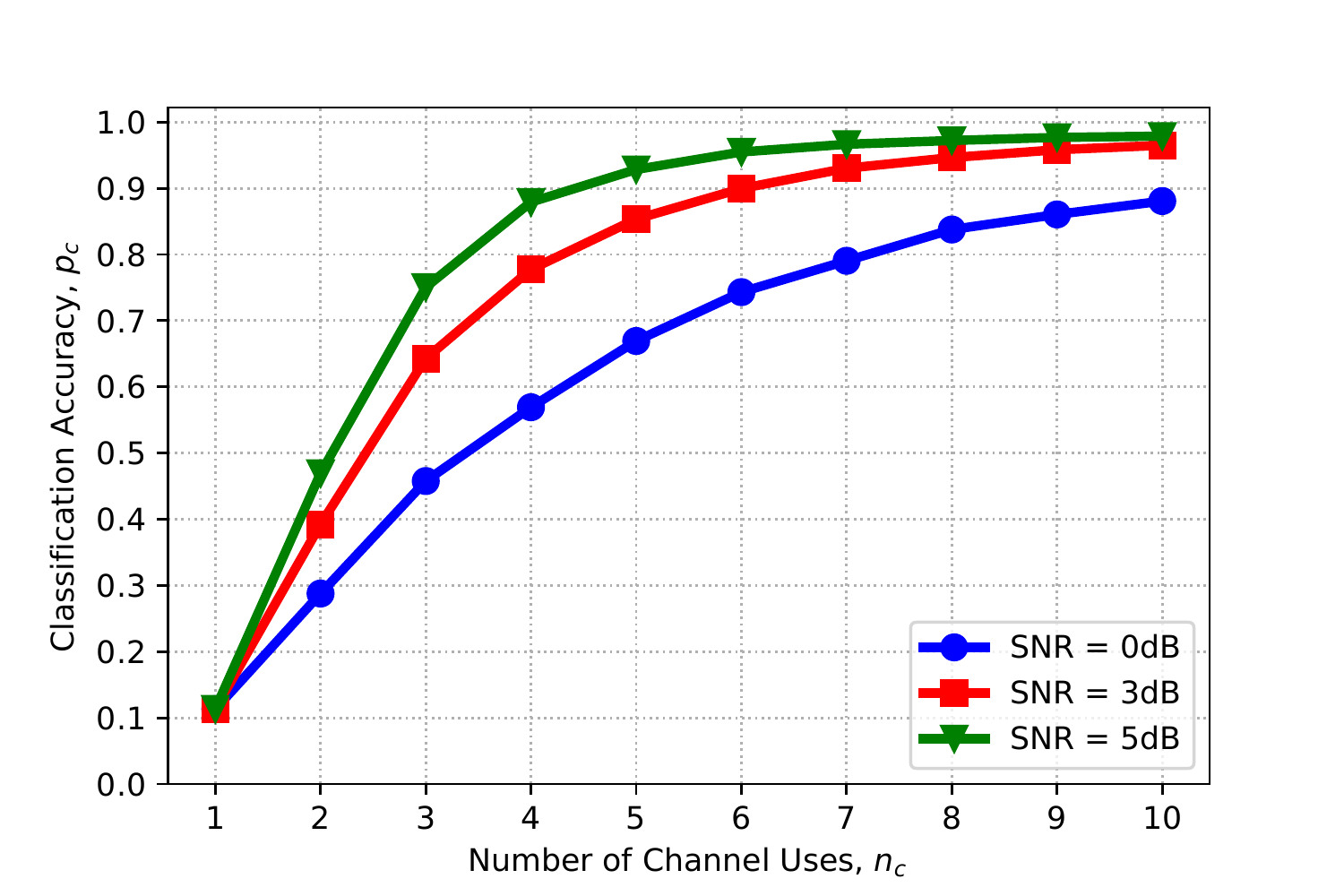}
\caption{Data: MNIST, Model: CNN.}
\label{fig:accuracyvsnchannelMNISTCNN}
\end{subfigure}
\begin{subfigure}[b]{0.32\textwidth}
\centering
\includegraphics[width=\columnwidth]{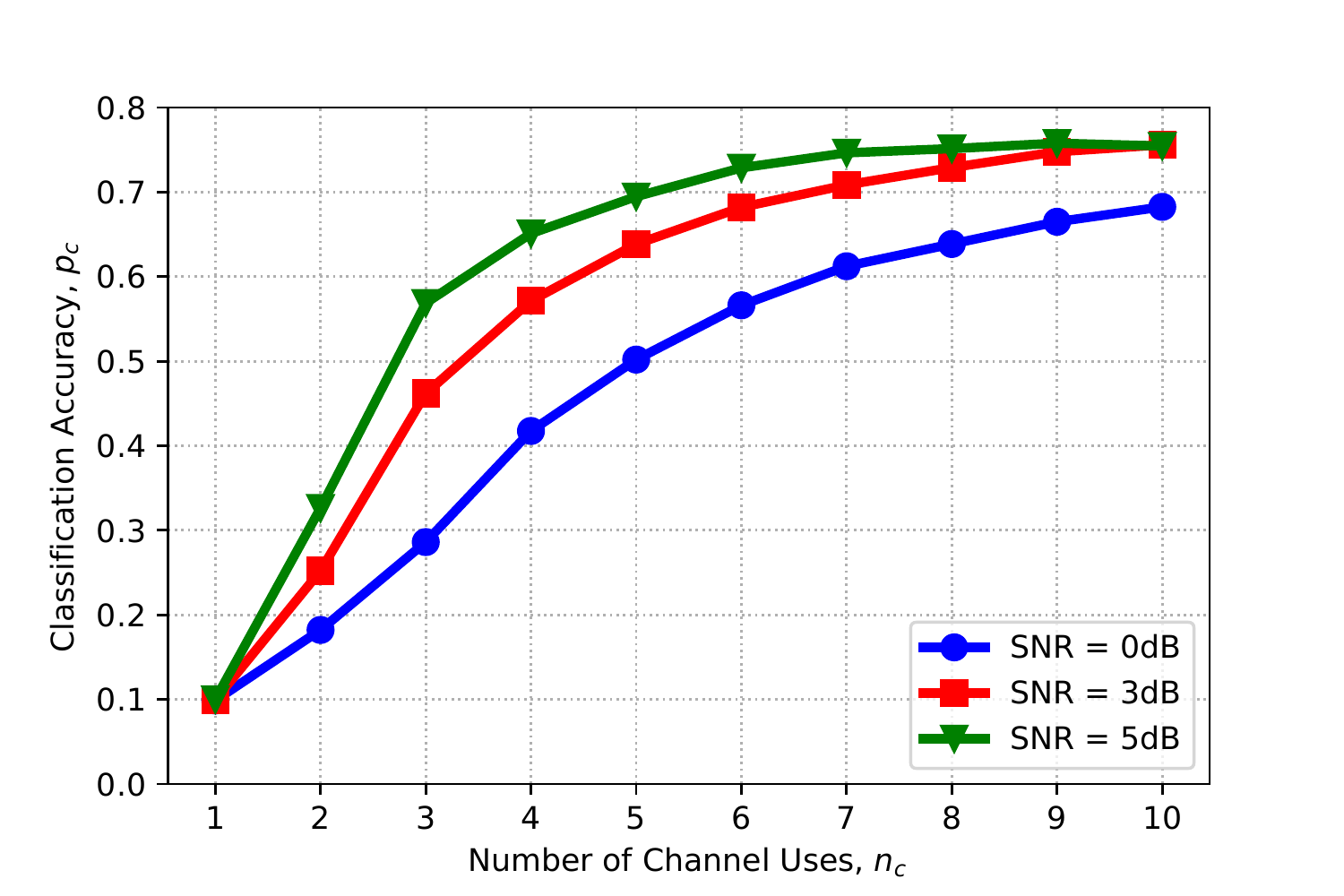}
\caption{Data: CIFAR-10, Model: CNN.}
\label{fig:accuracyvsnchannelCIFARTCNN}
\end{subfigure}
\caption{Classification (task) accuracy $p_c(n_c)$ as a function of the number of channel uses, $n_c$, for different SNR levels, datasets and model types.}
\label{fig:accuracy}
\end{figure*}

Fig.~\ref{fig:accuracy} shows the classification accuracy, $p_c$, namely the probability of correct classification (averaged over all labels), as a function of $n_c$. The accuracy $p_c$ is the highest for the MNIST data when the the more complex CNN model is used, whereas it is the lowest for the more difficult CIFAR-10 data even when the CNN is used (the FNN has poor performance for the CIFAR-10 dataset so it is not considered). For all data and model types considered, the accuracy $p_c$ increases, as $n_c$ increases. In the meantime, the service time, namely $n_c$, for the classification task to complete for each data sample also increases. This leads to the trade-off with freshness of task outcomes that we study next in Section \ref{sec:PAoTI}.

\section{Peak Age of Task Information in Task-Oriented Communications} \label{sec:PAoTI}
In this paper, we focus on the peak age of information (PAoI) \cite{PAoI} as a more tractable metric. We assume that the data samples (images) arrive at a First Come First Served (FCFS) queue of the transmitter according to a Poisson process (such that the interarrival time for data samples has exponential distribution). The service time (to classify each image) at the receiver is deterministic and given by $n_c$ (since each image is transmitted over $n_c$ channel uses). Thus, we consider an M/D/1 queue for the end-to-end operation of classification task from the arrival of images at the transmitter until they are classified at the receiver. We assume that the age is reduced only when correct information is obtained at the receiver, i.e., the classification outcome is correct. Otherwise, the age continues to increase linearly with time at unit rate. 

We denote by $t'_k$ the time instants at which the status update arrives (i.e., an image is received and classified) at the receiver.
At time instant $\hat{t}$, $N(\hat{t}) = \max\{k|t'_k \leq \hat{t}\}$ corresponds to the index of the most recently received update and $U(\hat{t}) = t_{N(\hat{t})}$ is the time stamp of the most recently received update. Then, the AoI is defined as $\Delta(t) = t - U(t)$ at time $t$.
The age $\Delta(t)$ reaches a peak $A_k = T_{k-1} + D_k$ or $Y_k + T_k$ the instant before the service completion at time $t'_k$, where 
 $T_k = t'_k - t_k$ is the time spent in the system, $D_k = t'_k - t'_{k-1}$ is the $k$th interdeparture time for data samples classified at the receiver, and $Y_k$ is the $k$th interarrival time for data samples at the transmitter. Assuming that the status updating system is stationary and ergodic, the PAoI is given by 
\begin{align} 
\Delta^{(\text{PAoI})} = E[A_k] &= E[T_{k-1}] + E[D_k] \nonumber \\
&=  E[Y_k] + E[T_k]. \label{eq:1}
\end{align}
Note that the PAoI is closely related to the AoI ($T_{k-1}$ and $A_k$ are sufficient to reconstruct the age process $\Delta(t)$). However, the PAoI does not require computing the joint expectation of interarrival time and systems time $E[T_n Y_n]$. Therefore, the PAoI is considered more tractable. 

Note that not every status update is successfully received, i.e., classification may yield an error, and then the age is not updated. Then, $T_k$ needs to be decomposed into two terms (similar to status updates over erasure channels in \cite{chen2016age}), the interarrival time for status updates that are successfully received, $\hat{T}_k$ and system time for 
the update that consists of time for waiting in queue, $W_k$, and time for service $S_k$. Note that $\hat{T}_k$ measures the interarrival time between the update $k$ and the successfully received update that has not arrived before the update $k$. Then, the PAoTI is given by
\begin{align} \label{eq:2}
\Delta^{(\text{PAoTI})} = E[Y_k] + E[\hat{T}_k] + E[W_k] + E[S_k]
\end{align}
For the M/D/1 queue, the terms $E[Y_k]$, $E[W_k]$, and $E[S_k]$ in (\ref{eq:2}) are given by
\begin{align}
E[Y_k] = \frac{1}{\lambda}, \: \: \:
E[W_k] = \frac{\rho}{2\mu (1-\rho)}, \: \: \:
E[S_k] = \frac{1}{\mu} \label{eq:5},
\end{align}
where $\lambda$ is the interarrival rate for status updates, $\rho = \lambda/\mu$ is the utilization ratio and $\mu = 1/n_c$ is the service rate. Next, we derive the term $E[\hat{T}_k]$ in (\ref{eq:2}). Let $\mathcal{E}_k$ and $\bar{\mathcal{E}}_k$ denote the events that the $k$th classification outcome is correct (and therefore, the age is reduced) and incorrect (and therefore, the age is not reduced), respectively. Then, $E[\hat{T}_k]$  in (\ref{eq:2}) is written as 
\begin{align} 
E[\hat{T}_k] =& P(\mathcal{E}_k) E[\hat{T}_k| \mathcal{E}_k]  \nonumber \\  
&+ (1-P(\mathcal{E}_k)) (E[T_k] + E[\hat{T}_{k+1}|\bar{\mathcal{E}}_k]), \label{eq:6}
\end{align}
where $P(\mathcal{E}_k)$ is the probability of correct classification of the signal corresponding to the $k$th data sample, and
\begin{align} 
E[\hat{T}_k| \mathcal{E}_k] &= 0, \label{eq:7} \\
E[\hat{T}_k| \bar{\mathcal{E}}_k] &= E[Y_k] + E[\hat{T}_{k+1}]. \label{eq:8}
\end{align}
Define $P(\mathcal{E}_k) = p_c$ as the probability of correct classification (assuming independent and identically distributed classification outcomes). $P(\mathcal{E}_k)$ depends on the type of data, and model, the SNR, and the number of channel uses $n_c$, as shown in Fig.~\ref{fig:accuracy}. As we focus on the trade-off driven by $n_c$, we express the probability $p_c$ as $p_c(n_c)$. From (\ref{eq:6})-(\ref{eq:8}), $E[\hat{T}_k]$ satisfies 
\begin{align} \label{eq:9}
E[\hat{T}_k] = \left(1-p_c(n_c) \right) \left( \frac{1}{\lambda} + E[\hat{T}_{k+1}] \right).
\end{align}
Considering $E[\hat{T}_{k}] = E[\hat{T}_{k+1}]$, we obtain $E[\hat{T}_{k}]$ from (\ref{eq:9}) as
\begin{align} \label{eq:10}
E[\hat{T}_k] = \frac{1-p_c(n_c)}{\lambda \: p_c(n_c)}.
\end{align}
Then, the average PAoTI, $\Delta^{(\text{PAoTI})}$, in (\ref{eq:2}) can be derived  from (\ref{eq:5}) and (\ref{eq:10}) as
\begin{equation} \label{eq:11}
\Delta^{(\text{PAoTI})} = \frac{1}{\lambda} + \frac{1-p_c(n_c)}{\lambda \: p_c(n_c) }+ \frac{\rho}{2\: \mu \: (1-\rho)} +  \frac{1}{\mu}.
\end{equation}
By setting $\mu = 1/n_c$ and $\rho = \lambda/\mu$ in  (\ref{eq:11}), $\Delta^{(\text{PAoTI})}$ is expressed as a function of $\lambda$, $n_c$ and $p_c(n_c)$ as 
\begin{equation} \label{eq:PAoTI}
   \Delta^{(\text{PAoTI})} = \frac{1}{\lambda \: p_c(n_c)}+ \frac{(2-\lambda \: n_c) \: n_c}{2\: (1-\lambda \: n_c)}.
\end{equation}

\section{Performance Evaluation and Design Guideline} \label{sec:perf}
\begin{figure*}[h!]
\centering 
\begin{subfigure}[b]{0.32\textwidth}
\centering
\includegraphics[width=\columnwidth]{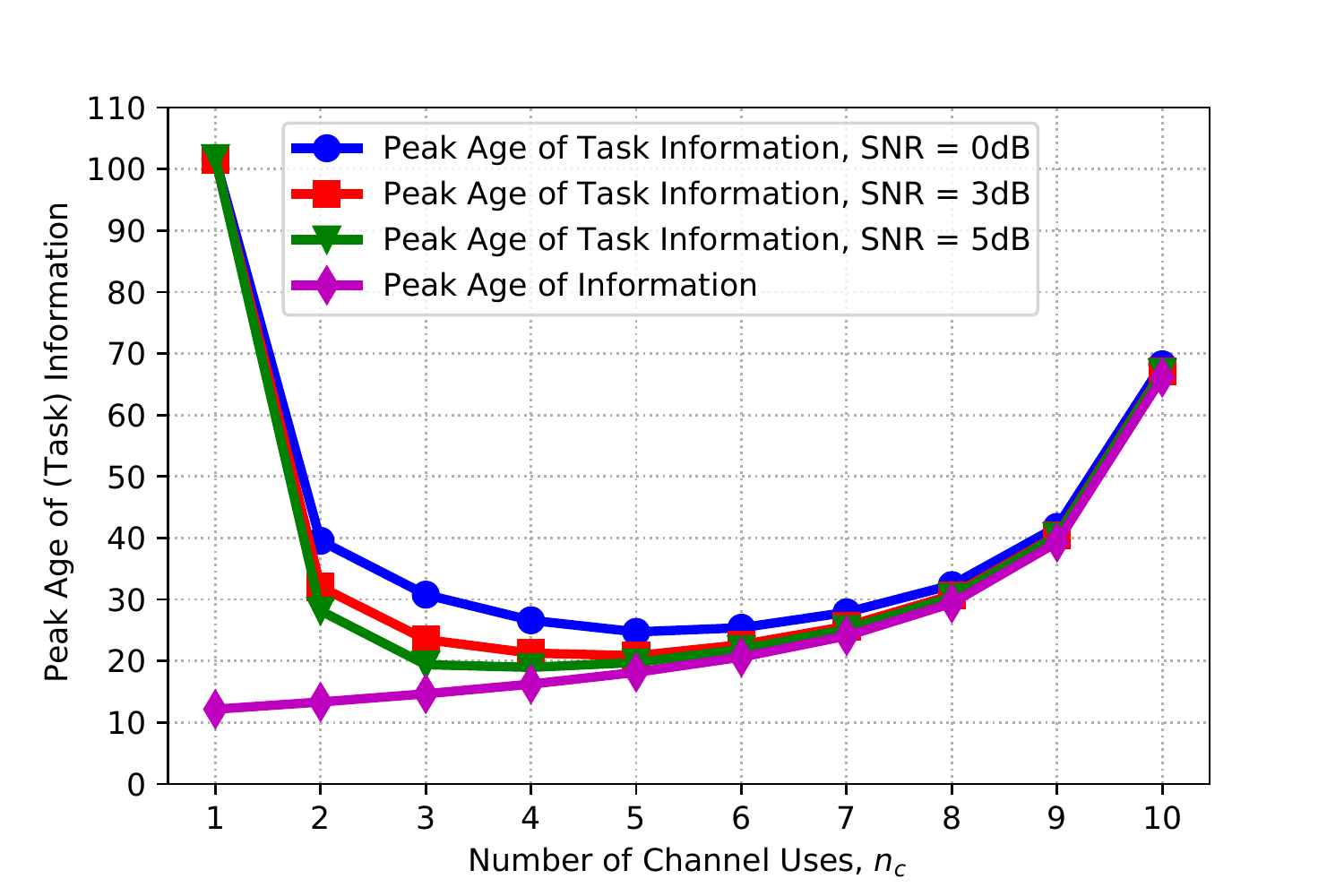}
\caption{Data: MNIST, Model: FNN.}
\label{fig:AOIMNISTFNN3}
\end{subfigure}
\begin{subfigure}[b]{0.32\textwidth}
\centering
\includegraphics[width=\columnwidth]{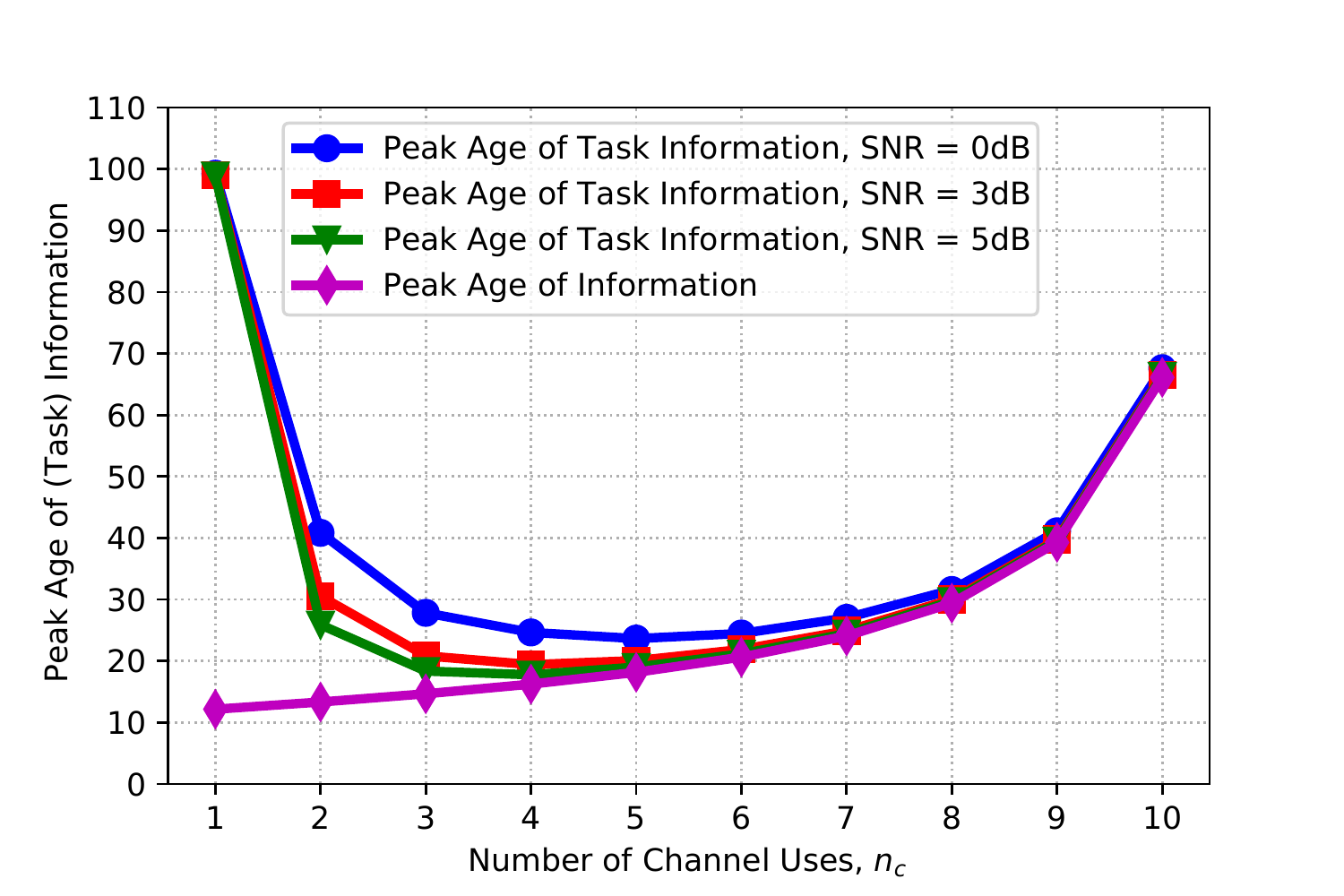}
\caption{Data: MNIST, Model: CNN.}
\label{fig:AOIMNISTCNN3}
\end{subfigure}
\begin{subfigure}[b]{0.32\textwidth}
\centering
\includegraphics[width=\columnwidth]{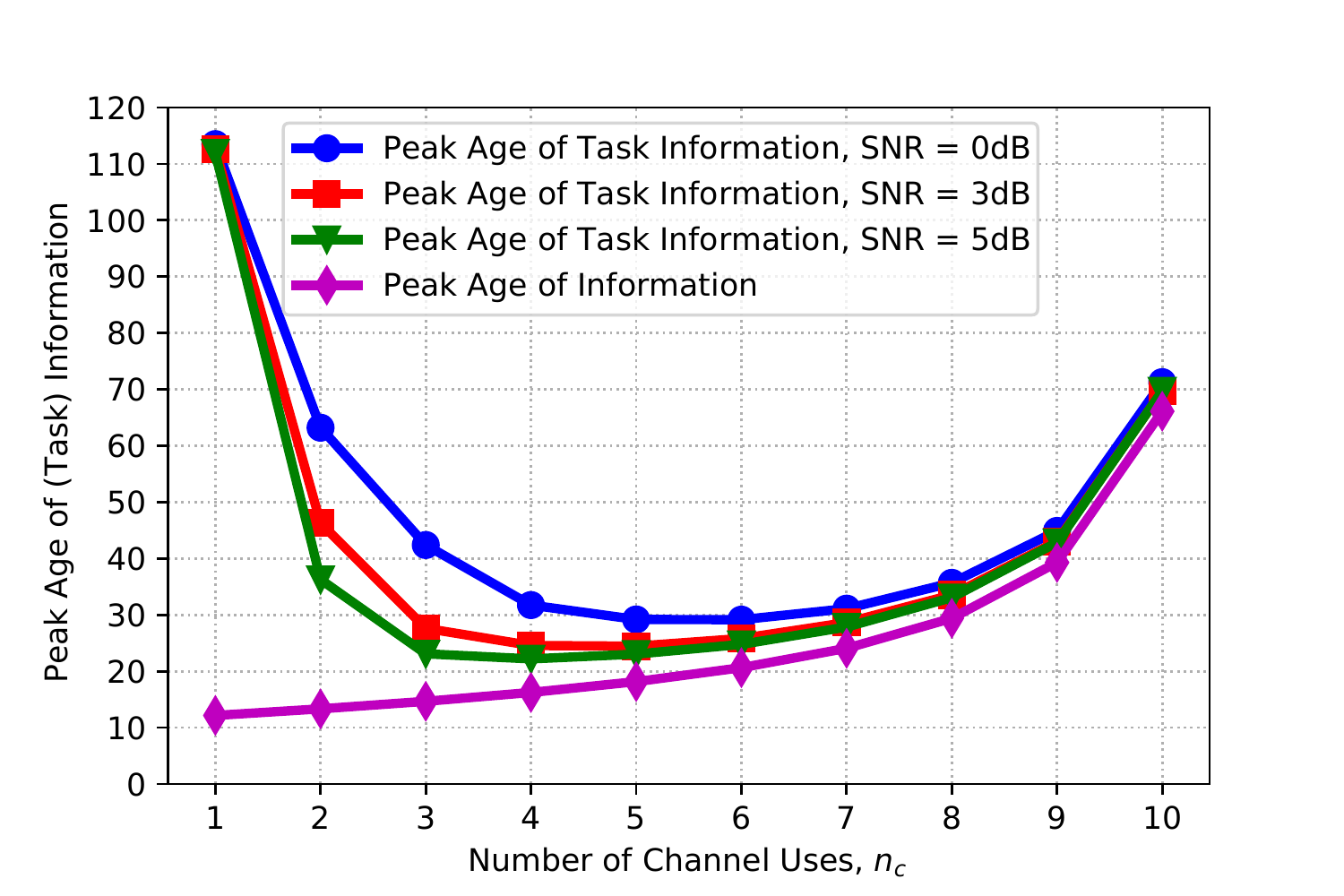}
\caption{Data: CIFAR-10, Model: CNN.}
\label{fig:AOICIFARCNN3}
\end{subfigure}
\caption{PAoTI as a function of the number of channel uses, $n_c$, for different SNR levels, datasets and model types, where the arrival rate, $\lambda$, is $0.09$.}
\label{fig:AOInc}
\vspace*{-0.2cm}
\end{figure*}

In this section, we start with the evaluation of the PAoTI as a function of the system parameters, namely the number of channel uses, $n_c$, the SNR of the channel, the arrival rate, $\lambda$, the data type (MNIST vs. CIFAR-10) and the model type (FNN vs. CNN). Fig.~\ref{fig:AOInc} shows the PAoTI as a function of the number of channel uses and compares it with the PAoI for different SNR levels, datasets and models types, where the arrival rate is $0.09$. The difference between the PAoTI and PAoI is that the age is reduced in PAoI whenever a classification takes place at the receiver, whereas the age is reduced in PAoTI only when the classification of a received signal at the receiver yields the correct label. 

The PAoI increases monotonically with $n_c$, since the service time is $n_c$. Also, the PAoI is independent of the classification accuracy, $p_c$, and it is not affected by the SNR as well as the dependence of $p_c$ on $n_c$. The PAoTI has a more complicated dependence on $n_c$ according to (\ref{eq:PAoTI}). When $n_c$ is small, the service time is also small and this is a driving factor to reduce the PAoTI. However, $p_c$ decreases rapidly with smaller $n_c$, as shown in Fig.~\ref{fig:accuracy}. Therefore, when $n_c$ is very small, the PAoTI first increases with $n_c$. However, as $n_c$ increases, the improvement of $p_c$ slows down, whereas the service rate decreases linearly with $n_c$. Therefore, the decrease in the PAoTI slows down when $n_c$ increases and the PAoTI eventually reaches its minimum for a moderate $n_c$. We denote this best $n_c$ as $n_c^*$. We compute $n_c^*$ as $5$, $4$ and $4$ for MNIST data and $6$, $5$ and $4$ for CIFAR-10 data when the SNR is $0$dB, $3$dB and $5$dB, respectively. Beyond $n_c^*$, $p_c$ starts to saturate with increasing $n_c$, whereas the service rate continues to decrease linearly with $n_c$. Thus, the PAoTI starts increasing with $n_c > n_c^*$. Since $p_c$ increases with the SNR, as shown in Fig.~\ref{fig:accuracy}, the PAoTI decreases with the SNR. Among the data types and model types, the smallest PAoTI is achieved for the MNIST data and the CNN model. On the other hand, $p_c$ is the smallest for the CIFAR-10 dataset for which the PAoTI is also the highest. The gap between the PAoTI and the PAoI increases when $n_c$ or the SNR decreases, i.e., when $p_c$ decreases and its negative effect on the PAoTI increases. 

\begin{figure*}[h]
\centering 
\begin{subfigure}[b]{0.32\textwidth}
\includegraphics[width=\columnwidth]{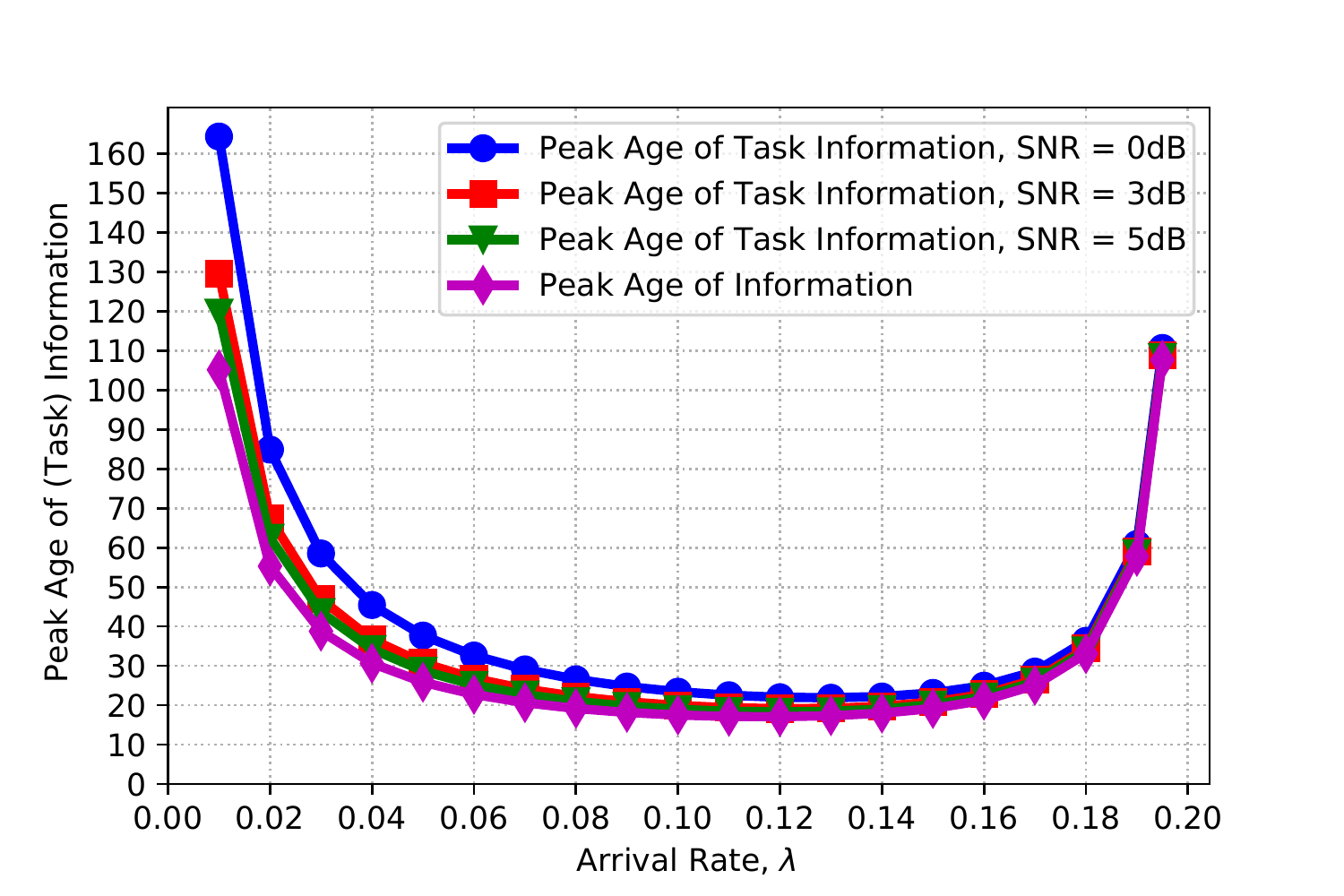}
\caption{Data: MNIST, Model: FNN.}
\label{fig:AOIMNISTFNN4}
\end{subfigure}
\begin{subfigure}[b]{0.32\textwidth}
\centering
\includegraphics[width=\columnwidth]{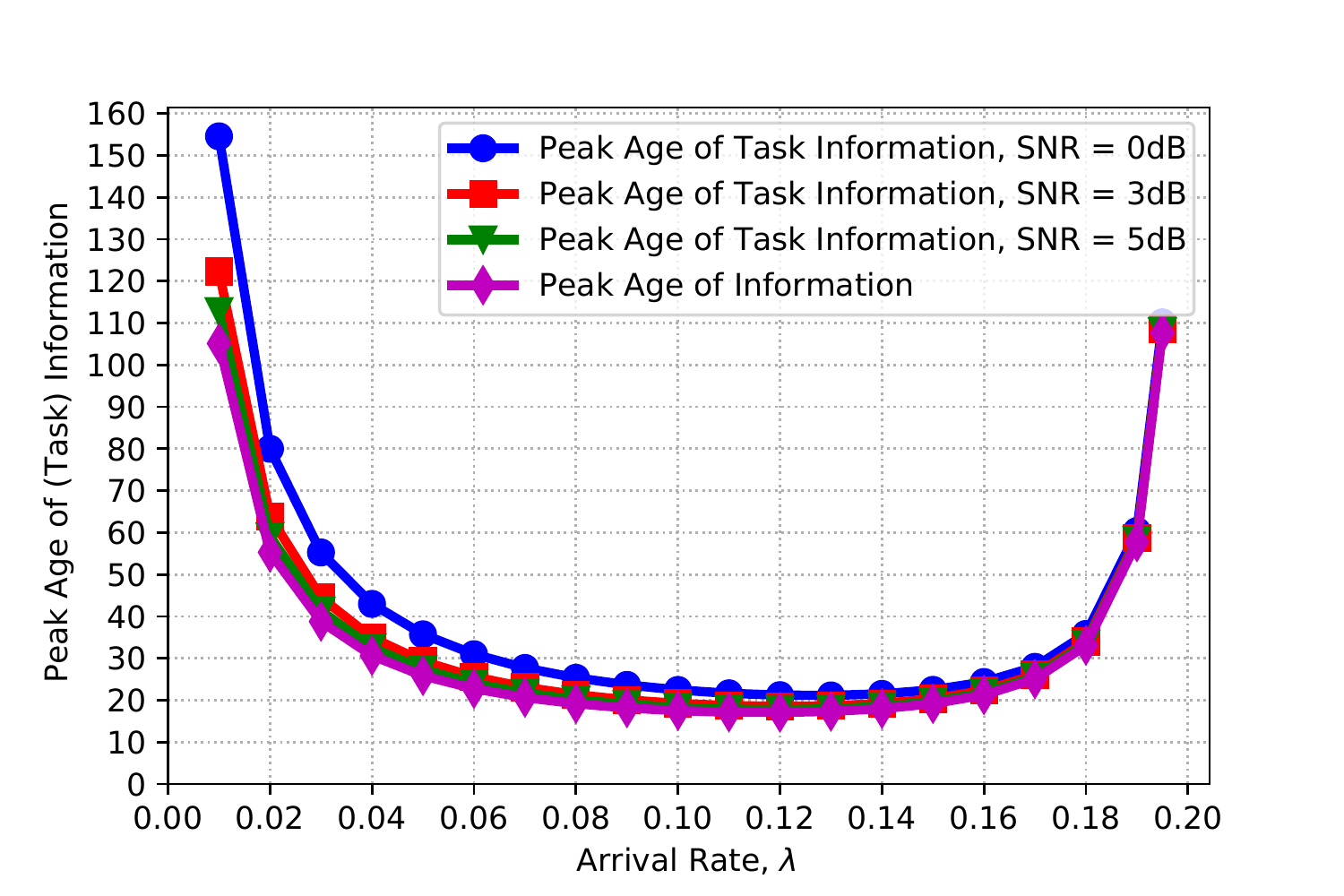}
\caption{Data: MNIST, Model: CNN.}
\label{fig:AOIMNISTCNN4}
\end{subfigure}
\begin{subfigure}[b]{0.32\textwidth}
\centering
\includegraphics[width=\columnwidth]{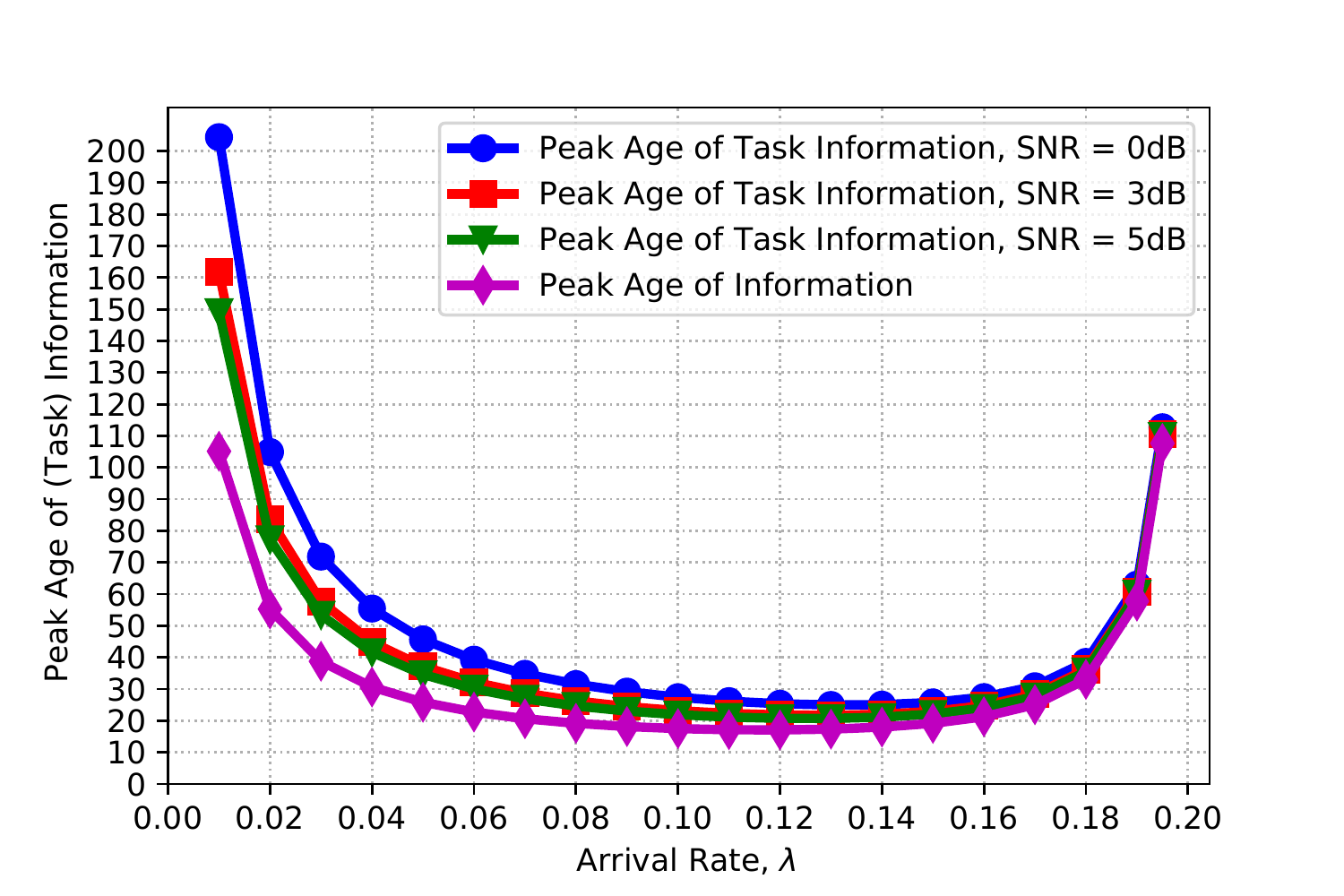}
\caption{Data: CIFAR-10, Model: CNN.}
\label{fig:AOICIFARCNN4}
\end{subfigure}
\caption{PAoTI as a function of the arrival rate, $\lambda$, for different SNR levels, datasets and model types, when the number of channel uses, $n_c$, is $5$.}
\label{fig:AOIna}
\vspace*{-0.4cm}
\end{figure*}

Next, we evaluate the effect of arrival rate, $\lambda$, on the PAoTI and the PAoI. Fig.~\ref{fig:AOIna} shows the PAoTI as a function of $\lambda$ and compares it with the PAoI for different SNR levels,  datasets and model types, when $n_c$ is $5$. Consistent with the known dependence of the age on the utilization, the PAoTI and PAoI first decrease as $\lambda$ increases (when the service rate is fixed), reach the minimum values, and then start increasing with $\lambda$. 

Since $n_c$ is a design parameter for the construction of the encoder-decoder pair in task-oriented communications, the PAoTI provides important design guidelines on the selection of $n_c$ for the freshness of task updates. The analysis so far assumes that the SNR and the arrival rate are known in advance such that an appropriate $n_c$ can be selected for a low level of the PAoTI. Next, we consider the case that the SNR and the arrival rate are unknown to the transmitter and the receiver, and they update $n_c$ (and the corresponding encoder and decoder models from Table~\ref{table:DNN}) based on the measured PAoTI over time. Let $n_c(t)$ denote the number of channel uses, $n_c$, that is adopted at time $t$. When a classification is performed (namely at any time instant $t'_k$), $n_c$ is updated as 
\begin{align}
n_c(t) = \left[n_c(t'_k) + \delta_k \right]^+, \quad \text{for } \: t > t'_k,
\end{align}
where $\delta_k$ is the change to $n_c$ at the $k$th update and $[x]^+ = \max(x,0)$. The update $\delta_k$ can be constructed as 
\begin{align}
\delta_k =
\begin{cases}
  \delta_{k-1}, & \text{if } \Delta^{(\text{PAoTI})}_{k} > \Delta^{(\text{PAoTI})}_{k-1}
\\
  - \delta_{k-1}, & \text{else if } \Delta^{(\text{PAoTI})}_{k} < \Delta^{(\text{PAoTI})}_{k-1} 
  \\
  \tilde\delta, & \text{otherwise},
\end{cases}
\end{align}
where $\delta_k \in \{-1, +1\}$ and $\tilde\delta\in \{-1, +1\}$ is a random variable of two-point distribution with values $-1$ and $+1$, each with probability $1/2$. Note that the exploration on $n_c$ can be limited by adding $0$ to the set of $\delta_k$ updates either by replacing $\tilde\delta$ or randomizing the $\delta_{k-1}$ versus $-\delta_{k-1}$ updates. 

\begin{figure}[h]
\centering
\includegraphics[width=0.85\columnwidth]{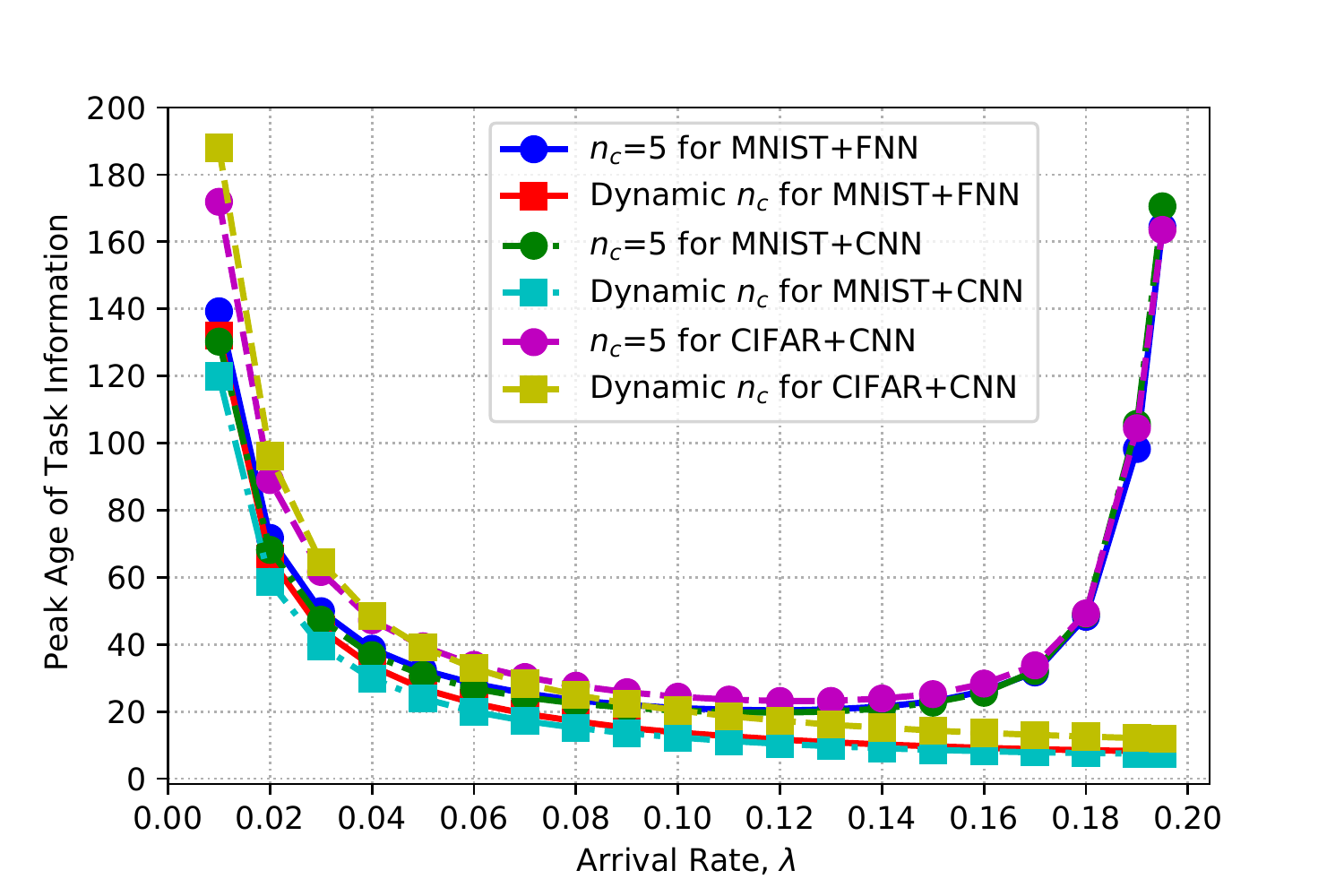}
 \caption{PAoTI (averaged over SNR levels) as a function of arrival rate, $\lambda$, for fixed number of channel uses ($n_c = 5$, which is the best choice for the arrival rate $\lambda = 0.09$) and dynamic number of channel uses.} \label{fig:AOIdynamic}
 \vspace*{-0.2cm}
\end{figure} 

Fig.~\ref{fig:AOIdynamic} shows the PAoTI for dynamic $n_c$ updates as a function of arrival rate, $\lambda$, for different SNR levels and compares it with the fixed $n_c$ case ($n_c = 5$). The PAoTI is measured and averaged over up to 200,000 status update events. Note that the dynamic scheme keeps the PAoTI very close to the fixed $n_c$ case for small $\lambda$ and reduces the PAoTI significantly when $\lambda$ is large such that the utilization significantly increases for the fixed $n_c$ case. Table~\ref{table:dynamic} shows the PAoTI that is measured and averaged for different arrival rates and SNRs. The dynamic $n_c$ reduces the PAoTI with respect to the fixed $n_c$ case by $47\%$ for MNSIT+FNN, $52\%$ for MNIST+CNN and $32\%$ for CIFAR-10+CNN. In other words, the PAoTI improvement is more when $p_c$ is higher, when the PAoTI is smaller and there is less randomness regarding when the age is reduced. 

\begin{table}[h!]
 \captionsetup{justification=centering}
 \caption{PAoTI averaged over SNR levels and arrival rates for fixed $n_c$ ($n_c = 5$) and dynamic update of $n_c$.}
	\label{table:dynamic}

	\centering
	{\vspace{0.2cm}
	\footnotesize
		\begin{tabular}{l|l|l|l}
& MNIST+FNN & MNIST+CNN & CIFAR-10+CNN  \\
\hline \hline
Fixed $n_c = 5$ & 46.35 & 45.74 & 52.48\\ \hline
Dynamic $n_c$ &  24.42 & 21.88 & 35.51
		\end{tabular}
	}
 \vspace*{-0.1cm}
\end{table}

\section{Conclusion} \label{sec:Conclusion}
We have studied the notion of age in task-oriented communications, where the goal of communications is to facilitate task execution, e.g., image classification, at the receiver by using data samples available at the transmitter. An encoder at the transmitter compresses data samples, and feature vectors of a small dimension are transmitted over the wireless channel with a small number of channel uses. The decoder at the receiver classifies the received signals instead of reconstructing data samples. This encoder-decoder pair is jointly trained by accounting for channel effects. Using MNIST and CIFAR-10 data with FNN or CNN models, we have assessed the increase in classifier accuracy and service time with the number of channel uses. We have introduced the concept of the PAoTI that measures the peak age in task-oriented communications, where the age increases with time unless a data sample arriving at the transmitter queue is classified correctly at the receiver. We have characterized the PAoTI as a function of the number of channel uses that emerges as a design feature for task-oriented communications. First, we have shown how to select the number of channel uses for the given arrival rate and SNR. Then, we have presented a dynamic scheme of updating the number of channel uses to reduce the PAoTI without knowing the arrival rate and the SNR in advance. Our approach captures the accuracy-latency trade-offs via the notion of PAoTI and identifies design mechanisms for task-oriented communications. 

\bibliographystyle{IEEEtran}
\bibliography{references}

\end{document}